\documentclass[twocolumn]{aastex631}

\newcommand\aastex{AAS\TeX}
\newcommand\latex{La\TeX}
\usepackage{graphicx}	% Including figure files
\usepackage{amsmath}	% Advanced maths commands

\begin{document}

\title[Role of Halo Angular Momentum Distribution Discontinuity]{Importance of Initial Condition on Bar Secular Evolution: Role of Halo Angular Momentum Distribution Discontinuity}

\correspondingauthor{Sandeep Kumar Kataria, Juntai Shen}
\email{skkataria.iit@gmail.com, jtshen@sjtu.edu.cn}

\author[0000-0003-3657-0200]{Sandeep Kumar Kataria}
\affiliation{Department of Astronomy, School of Physics and Astronomy, Shanghai Jiao Tong University, 800 Dongchuan Road, Shanghai 200240, China}
\affiliation{Key Laboratory for Particle Astrophysics and Cosmology (MOE) / Shanghai Key Laboratory for Particle Physics and Cosmology, Shanghai 200240, China
}

\author[ 0000-0001-5604-1643]{Juntai Shen}
\affiliation{Department of Astronomy, School of Physics and Astronomy, Shanghai Jiao Tong University, 800 Dongchuan Road, Shanghai 200240, China}
\affiliation{Key Laboratory for Particle Astrophysics and Cosmology (MOE) / Shanghai Key Laboratory for Particle Physics and Cosmology, Shanghai 200240, China
}

%% Note that the \and command from previous versions of AASTeX is now
%% depreciated in this version as it is no longer necessary. AASTeX 
%% automatically takes care of all commas and "and"s between authors names.

%% AASTeX 6.31 has the new \collaboration and \nocollaboration commands to
%% provide the collaboration status of a group of authors. These commands 
%% can be used either before or after the list of corresponding authors. The
%% argument for \collaboration is the collaboration identifier. Authors are
%% encouraged to surround collaboration identifiers with ()s. The 
%% \nocollaboration command takes no argument and exists to indicate that
%% the nearby authors are not part of surrounding collaborations.

%% Mark off the abstract in the ``abstract'' environment. 
\begin{abstract}
The dark matter halo properties, for example, mass, spin and concentration play a significant role in the formation and evolution of bars in disk galaxies. This study highlights the importance of a new parameter: the dark matter halo angular momentum distribution in the disk's central region. We experiment with N-body galaxy models having a disk and dark matter similar to Milky Way-type galaxies. In these models, we vary the discontinuity of the angular momentum distribution of the halo (the total spin is the same for all models). Our N-body experiments suggest that bar forms in all models after a few Gyr of disk evolution. However, in the secular evolution of the bar, as we evolve these models until 9.78 Gyr, the bar gains its strength in the model with the most continuous halo angular momentum distribution, and the bar loses strength for the most discontinuous halo angular momentum distribution. The secular evolution of the bar suggests that box/peanut/x-shaped bulges similar to those found in the Milky Way disk should be more pronounced in halos with continuous halo angular momentum distributions. This study demonstrates the importance of the initial condition setup of galaxy systems, namely the discontinuity in the dark matter halo angular momentum distribution for a given density distribution,  on the bar secular evolution in the disk galaxy simulations. Further, this study helps reconcile the conflicting results of bar secular evolution in a high-spinning halo of the recent literature. 

\end{abstract}

%% Keywords should appear after the \end{abstract} command. 
%% The AAS Journals now uses Unified Astronomy Thesaurus concepts:
%% https://astrothesaurus.org
%% You will be asked to selected these concepts during the submission process
%% but this old "keyword" functionality is maintained in case authors want
%% to include these concepts in their preprints.

%% From the front matter, we move on to the body of the paper.
%% Sections are demarcated by \section and \subsection, respectively.
%% Observe the use of the LaTeX \label
%% command after the \subsection to give a symbolic KEY to the
%% subsection for cross-referencing in a \ref command.
%% You can use LaTeX's \ref and \label commands to keep track of
%% cross-references to sections, equations, tables, and figures.
%% That way, if you change the order of any elements, LaTeX will
%% automatically renumber them.
%%
%% We recommend that authors also use the natbib \citep
%% and \citet commands to identify citations.  The citations are
%% tied to the reference list via symbolic KEYs. The KEY corresponds
%% to the KEY in the \bibitem in the reference list below. 

\section{Introduction} \label{sec:intro}
Bars are ubiquitous in disk galaxies, given two-thirds of spiral galaxies in the local universe possess bars \citep{Eskridge.et.al.2000,Marinova_2007,Delmestre.et.al.2007,Erwin.2018}. Observations show a constant bar fraction until redshift $z \approx0.84$ for high mass galaxies and constant up to $z\approx0.3$ for low mass galaxies \citep{Sheth.et.al.2008}, while other studies \citep{Jogee.etal.2004} claim that the bar fraction remains constant until $z\approx1$ irrespective of galaxy mass. Recent studies \citep{Guo.et.al.2023, Le.Conte.2023} quantified bars at high redshift ($z>1$) in near-infrared wavelengths with the high sensitivity of the JWST. Therefore, the bar plays a key role in the evolution of disk galaxies, given that bars survive in galaxies for a long time, around $t\approx6-7 $ Gyr. The recent cosmological simulations \citep{Rosas-Guevara.et.al.2020, Yetli.et.al.2022, Ansar.et.al.2023_Fire, Ansar.et.al.2023,Kataria.Vivek.2024,Fragkoudi.et.al.2024} also show that strong bars are mostly formed at higher redshifts ($z\approx0.5-1.5$).  
\\

Dynamically cold disks with the most rotation motion are prone to bar-type instabilities \citep{Ostriker.Peebles.1973,Saha.elmegreen.2018, Kataria.Das.2018,Kataria.Das.2019, Kataria_etal_2020}. Further, the bar undergoes vertical instabilities or vertical resonances, leading to boxy/peanut/x-shapes shapes \citep{Pfenniger.Friedli.1991, Raha.et.al.1991,Sellwood.Merritt.1994, Athanassoula.2003, Sellwood.Gerhard.2020, Li.et.al.2023}. Several studies have shown that apart from the visible matter in the disk, the properties of the dark matter halo also play an important role in bar formation and evolution. For example, \cite{Athanassoulla.2002} shows that dark matter halo with high concentration leads to stronger and thinner bars compared to less concentrated ones \citep{Athanassoulla.2002}. Triaxial halos promote bar formation although they suppress the bar secular evolution \citep{Berentzen.et.al.2006,Athanassoula.et.al.2013} and prolate halos massively thicken the bar by enhancing buckling events \citep{KumarA.et.al.2022}. The halo spin also promotes the bar formation \citep{Kanak.Saha.Naab.2013,Longetal.2014}. Some studies show that secularly evolved bar weakens with prograde halo spin as a result of enhanced buckling \citep{Longetal.2014,Collieretal.2018,Collieretal.2019} while other studies show that bar strength does not get affected with prograde halo spin \citep{Kataria.Shen.2022}. These studies indicate that dark matter halo plays a significant role in the dynamics of barred galaxies and needs to be understood better in its effects on bar evolution.
\\
\\
The surrounding dark matter halo properties can affect the angular momentum transfer from the disk to the halo and, therefore, the bar formation process \citep{Athanassoula.2003}. \cite{Weinberg.1985} shows in his work using analytical perturbation theory and semi-restricted N-body simulation that the bar experiences dynamical friction due to the surrounding halo. The importance of live halo was first highlighted by \cite{Sellwood.1980} before this; most work only included rigid dark matter halo. The bar experiences dynamical friction and slows down. Further, \cite{Debattista.Sellwood.2000} show that a significant amount of angular momentum is required in the dark matter halo to prevent the bar from slowing down. This study includes halos having a continuous distribution of halo angular momentum in the central region. The previous studies \citep{Collieretal.2019,Li.et.al.2023_haloS,joshi.widrow.2023} introduced the discontinuity in the distribution of halo's angular momentum by simply flipping a 
 constant fraction of the dark matter particle's  orbital velocities to generate high-spinning halos. This method has been incorporated routinely in N-body simulations without considering the possible consequences on the equilibrium of halo and non-axisymmetric instabilities in the disk. The role of the discontinuity of halo angular momentum distribution has not been fully explored yet, which motivates us to study this important subject.

 To obtain analytical insight on the disk instability due to the discontinuity in the distribution of halo's angular momentum, we perform a linear approximation analysis. The collisionless Boltzmann equation (CBE) describes the dynamics of a self-gravitating system,
 \begin{equation}
      \dfrac{\partial f}{\partial t} +\textbf{v}.\dfrac{\partial f}{\partial \textbf{x} }-\dfrac{\partial \Phi}{\partial \textbf{x}}.\dfrac{\partial f}{\partial \textbf{v}}=0,
 \end{equation}

along with the  Poisson's equation, 
\begin{equation}
    \nabla ^2 \Phi (\textbf{x}, t)= 4 \pi G \int d^3 \textbf{v} f(\textbf{x,v}
,t), \end{equation}

where $\Phi (\textbf{x},t)$ is the total potential and the distribution function (DF) f(\textbf{x,v},t) is the mass density in phase space. 

The following conditions provide that the isolated system is in equilibrium,
\begin{equation}
    \textbf{v}.\dfrac{\partial f_0}{\partial \textbf{x} }-\dfrac{\partial \Phi_0}{\partial \textbf{x}}.\dfrac{\partial f_0}{\partial \textbf{v}}=0 \hspace{3 mm} ; \hspace{3 mm} \nabla^2 \Phi_0=4 \pi G \int d^3 \textbf{v} f_0, 
\end{equation}
where the equilibrium distribution $f_0(\textbf{x,v})$ and  potential $\Phi_0$ are time-independent.

The equilibrium system is now subject to a weak gravitational perturbation due to external potential $\epsilon \Phi_1 (\textbf{x},t)$ where $\epsilon \ll 1$. The DF and potential response of the system are given by
\begin{equation}
    f(\textbf{x,v},t)=f_0(\textbf{x,v}) +  \epsilon f_1(\textbf{x,v},t) ,
\end{equation}  

\begin{equation}  
    \Phi(\textbf{x},t) = \Phi_0(\textbf{x},t) + \epsilon\Phi_1(\textbf{x},t)
\end{equation}

Now following the section 5.1.2 of \citep{BT.2008}, the solution to linearized collisionless Boltzmann equation (CBE) is given by;
\begin{equation}\label{f1_growth}
    f_1(\textbf{x,v},t)= \int_{- \infty}^{t} dt'\bigg(\dfrac{\partial f_0}{\partial \textbf{v}}.\dfrac{\partial \Phi_1}{\partial \textbf{x}}-\dfrac{\partial f_0}{\partial \textbf{x}}.\dfrac{\partial \Phi_1}{\partial \textbf{v}}\bigg) ,
\end{equation}

here the partial differentials are calculated along unperturbed orbits $\textbf{x}(t'), \textbf{v} (t')$ that goes to $\textbf{x, v}$ at time t. The second term is zero given that $\Phi_1$ does not depend on velocity. Therefore we can write the following equation.

\begin{equation} \label{f1_growth2}
    f_1(\textbf{x,v},t)= \int_{- \infty}^{t} dt'\bigg(\dfrac{\partial f_0}{\partial \textbf{v}}.\dfrac{\partial \Phi_1}{\partial \textbf{x}}\bigg) ,
\end{equation}

 \begin{figure*}
    \centering
    \includegraphics[scale=0.43]{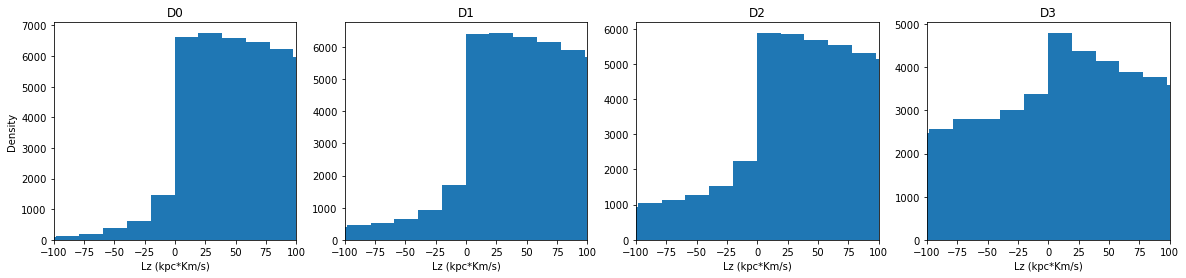}
    \caption{ The halo angular momentum distribution discontinuity about $L_z=0$ decreases from models D0 to D3.}
    \label{fig:AM_discon}
\end{figure*}

As we are interested in the growth of $f_1(\textbf{x,v},t)$ with time in a system where $f_0 (\textbf{x,v})$ is discontinuous concerning angular momentum or velocity distribution. In this scenario, $f_1 (\textbf{x,v},t)$ can grow to arbitrarily large values, given that $\dfrac{\partial f_0}{\partial \textbf{v}}$  diverges at zero tangential velocity. The conditions under which the linear perturbation in the distribution function  $f_1(\textbf{x,v},t)$ doesn't diverge include (1) perturbing force  $\dfrac{\partial \Phi_1}{\partial \textbf{x}}$ is close to zero. (2) If $\dfrac{\partial \Phi_1}{\partial \textbf{x}}$ is orthogonal to $\dfrac{\partial f_0}{\partial \textbf{v}}$ then $\Phi_1$ must be a purely axisymmetric perturbation.   
 Of course, the linear analysis is overly simplified. To understand the non-linear regime, we motivate N-body simulations of such systems for further probe.

 Further, we aim to investigate the role of halo angular momentum distribution discontinuity in answering questions like (1) How does the discontinuity of the angular momentum distribution in the central regions of the halo affect the bar formation process? 2) How does the secular growth of the bar get affected? (3) How does the buckling of the bar get affected? (4) How do the box/peanut/x-shaped bulges get affected?

In this article, we highlight the role of halo angular momentum discontinuity in bar formation and evolution processes for the first time. We also provide a broad message to the N-body community about the severe effect of the unnatural initial condition of the halo in the galaxy system. The content in the article is arranged in the following manner. In section \ref{method}, we discuss the underlying numerical method during this study. We elaborate on the main results in section \ref{Results} and the following discussion in section \ref{Discussion}. Finally, we summarize our results in section \ref{Conclusion}.

\section{Numerical Method}\label{method}

We have generated Milky-Way type initial galaxy models using Action-based-GAlaxy-Modeling Architecture (AGAMA) code \citep{Vasiliev.2019}. The code generates the equilibrium distribution of particles (position and velocity) for given components, namely, disk and dark matter halo, for all of our models. The code obtains the self-consistent galaxy model having input as the density and distribution function of all the components in the galaxy. The code solves a non-linearly coupled system of three linear integro-differential equations: potential from density using the Poisson equation and density from the supplied distribution function mapped with computed potential until the density converges.

The following Dehnen profile gives the density distribution for the dark matter halo.
\begin{equation}
    \rho_{\text{dm}}=\dfrac{M(3-\gamma)}{4 \pi  q a^3}\Bigg(\dfrac{r}{a}\Bigg)^{-\gamma}\Bigg(1+\dfrac{r}{a}\Bigg)^{(\gamma-4)}
\end{equation}

where $M$ is the total mass of the dark matter halo, $\gamma$ is the inner power-law slope, $a$ is the scale radius of the halo, and $q$ is the flattening parameter. For our model $M=57.42 \times 10^{10} M_{\odot}$, $a=14.32 $ kpc, $\gamma=1$ and $q=1$. The virial radius of the halo for all of our models is 140 kpc.

The following function gives the density of the disk.
\begin{equation}
    \rho_d=\Sigma_0 e^{-R/R_d}\times\dfrac{1}{4|h|}\text{sech}^2{\Big|\dfrac{z}{2h}\Big|}
\end{equation}

where $\Sigma_0$ is the surface density, $R_d$ is the scale radius and $h$ is the scale height. The values of these quantities are $\Sigma_0=1.2 \times 10^9 M_{\odot}/$kpc$^2$, $R_d=2.9$ kpc and $h=0.32$ kpc. The total mass in the disk component is 10$\%$ of the total halo mass. In our models, there is no stellar particle beyond the radius 15$R_d$. We use a million particles each for the disk and halo components of all the models.

The distribution function of the spherical dark matter component is `quasi-spherical', while for the disk component, it is `quasi-isothermal'. The halo is non-rotating, having an equal number of prograde and retrograde particles for this model generated using AGAMA. The Jeans theorem \citep{Jeans.1919J} states that reversing a particle's direction of motion for a given system in equilibrium will also be a solution to the collisionless Boltzmann equation. We reverse the $v_x$ and $v_y$ components of velocities of the dark matter halo particles in the central region to increase the halo spin and angular momentum distribution discontinuity around $L_z=0$.  Here spin is defined as \citep{Bullock.et.al.2001}:
    
    \begin{equation}
         \lambda=\dfrac{J}{\sqrt{2GMR}}  
         \label{eq:lambda}
    \end{equation}
       
Here $J$ is the magnitude of the specific angular momentum of the halo, $M$ is the mass of the halo within the virial radius, and $R$ is the virial radius of the halo.  

We have assigned a random probability ($x$) between 0 and 1 for all the halo particles with negative angular momentum ($L_z$) using a uniformly distributed random number generator. Afterwards, 
we have reversed the $v_x$ and $v_y$ velocity components of particles having probabilities ($x$) higher than the following probability function which depends on the angular momentum of particles \citep{Chiba.Kataria.2024} and inspired from \cite{Debattista.Sellwood.2000}.
\begin{equation}
g(L_z)=F \tanh\Bigg(\chi \dfrac{L_z}{L_s}\Bigg)
\label{prob}
\end{equation}

Here $F$ and $\chi$ control the step size and sharpness of the halo angular momentum discontinuity respectively. $L_z$ is the angular momentum of individual halo particle which is normalized by a constant $L_s$, in our case it is the maximum angular momentum of a particle in the dark matter halo distribution. We generate four models named D0, D1, D2, and D3, with a spin value ($\lambda$) of 0.1. We have used $F$ values to be 1, 0.9, 0.7 and 0.18 respectively from D0 to D3 models such that the step size of the halo angular momentum distribution discontinuity decreases. We have used $\chi$ values to be 100 for models D0, D1, D2 and 1 million for D3. In order to achieve spin ($\lambda$) of value 0.1 we reverse the halo particles (having $x>g(L_z)$) with the negative angular momentum larger than $-$3800, $-$5500 and $-$9000 for model D0, D1 and D2 respectively. While for the model D3 reverse all the particles with negative angular momentum.

\begin{figure}
    \centering
    \includegraphics[scale=0.55]{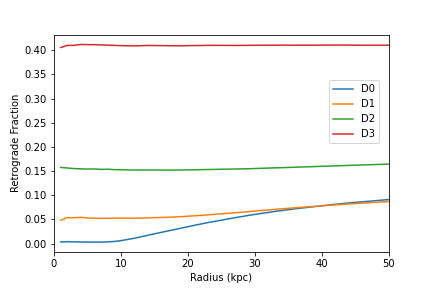}
    \caption{Radial variation of the retrograde fraction of dark matter halo orbits in all the galaxy models.}
    \label{fig:retro}
\end{figure}

We have tested whether these uppercut in $L_z$ reversal affect our results. We show that the simulation of model D0 by reversing particles with an uppercut and without upper cut results in a similar evolution of galaxy for which we discuss the details in appendix \ref{appendix}. The upper limit on angular momentum is introduced such that the spin of dark matter halo is equal to 0.1 which results in higher values without imposing the upper limit.    
The reversal of halo particle orbits in the above-mentioned rule results in a decrease of halo angular momentum distribution discontinuity around $L_z=0$ from model D0 to D3, as seen in Figure \ref{fig:AM_discon}. Figure \ref{fig:retro} shows the reversal also leads to the radial variation in retrograde orbit fraction. The fraction of reversed orbits at a given radius $R$ is given by
$f_{\text{reversed}}=0.5- f_{\text{retrograde}}$. 

%We have reversed the particle direction within the 80 kpc region having various energy ranges. These energy ranges in code units for models D0, D1, D2, and D3 are ($-$250000, 0), ($-$200000,$-$38000), ($-$150000,$-$36000) and ($-$65000,$-$12000), respectively. There are more particles in the higher energy end compared to the lower energy end which contains very few particles (the differential energy distribution $N(E)$ is an increasing function of $E$ as shown in Figure \ref{fig:Energy_dist}). The lower energy range particles having the most negative energy are the most bound in the central region. % 

We have evolved the galaxy models using the GADGET-2 code \citep{Springel.2005} until 9.78 Gyr. The angular momentum and energy conservation are within 0.1$\%$ in all the simulations. The softening length for the halo and disk components have been chosen as 30 and 25 pc, respectively. We set the opening angle of the tree algorithm to be 0.4. Throughout the paper, we describe our results in terms of code units i.e. unit mass equal to $10^{10}$  M$_{\sun}$, unit velocity is 1 km/s and unit length is 1 kpc.

\section{Results} \label{Results}

\begin{figure*}
    \centering
    \includegraphics[scale=0.65]{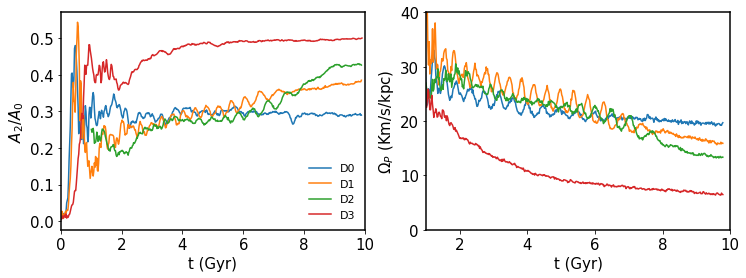}
    \caption{Left panel shows the evolution of bar strength with time. Model D3 shows the secular growth of bar strength, while bar strength dampens for model D0. The right panel shows the pattern speed evolution with time. The decrease in pattern speed for model D3 is faster compared to model D0 }
    \label{fig:BS_PS}
\end{figure*}
There are various ways bar strength has been defined in the literature \citep{Combes.Sanders.1981,Athanassoula.2003}. In this study, we look for stellar contribution to $m$=2 Fourier mode as a proxy for bar strength.

\begin{equation}
a_2(R)=\sum_{i=1}^{N}  m_i \cos(2\phi_i)\\ \hspace{0.5cm}
b_2(R)=\sum_{i=1}^{N} m_i \sin(2 \phi_i)
\label{equation:FM}
\end{equation}

where $a_2$ and $b_2$ are calculated for all the disk particles, $m_i$ is mass of $i^{th}$ star, $\phi_i$ is the azimuthal angle. We have defined the bar strength as  
\begin{equation}
\frac{A_2}{A_0}= \frac{\sqrt{a_2 ^2 +b_2 ^2}}{\sum_{i=1}^{N} m_i} 
\label{eq:barstrength}
\end{equation}
So, the bar strength is derived using the cumulative value of $m=2$ Fourier mode for the whole disk, which also varies with radius.

The left panel of Figure \ref{fig:BS_PS} shows the evolution of bar strength in our models with increasing continuity of dark matter halo angular momentum distribution. We notice that bar instability is triggered within the first Gyr of evolution for all the models. The bar growth in secular evolution depends on the central angular momentum distribution of the surrounding dark matter halo. The bar strength grows in model D3, where we have a continuous distribution of halo angular momentum. As the discontinuity of angular momentum distribution increases, we see that the secular growth of the bar is suppressed. The final bar strengths of models are inversely correlated with the discontinuity of halo angular momentum distribution. We find the weakest bar in model D0 which has the highest discontinuity in halo angular momentum distribution.

The right panel of Figure \ref{fig:BS_PS} shows the evolution of pattern speed with time for all the models. We can see that the decrease in pattern speed for model D3 is the largest, which resonates with the largest secular growth of the bar. The almost non-varying pattern speed correlates with the almost constant bar strength of model D0 in the secular evolution. Unlike bar strength, final pattern speed directly correlates with the discontinuity of halo angular momentum distribution.

\begin{figure*}
    \centering
    \includegraphics[scale=0.65]{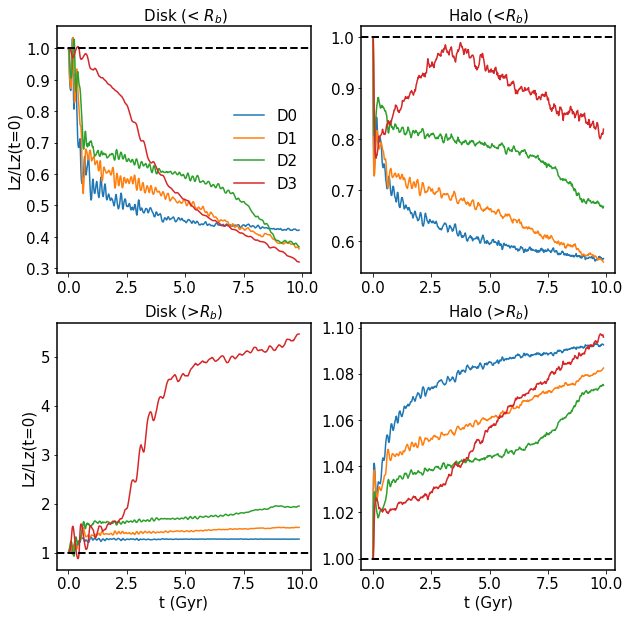}
    \caption{The time evolution of the angular momentum of disk and halo particles within bar length ($R_b$) region and outside the bar length ($R_b$) region for all four models. The horizontal dashed line corresponds to the initial angular momentum of the components.}
    \label{fig:AM_exchange}
\end{figure*}

Figure \ref{fig:AM_exchange} shows the angular momentum exchange in the disk within the bar region (radius $<$ bar length) and outside the bar region (radius $>$ bar length). We measure the bar length ($R_b$) as the radius where the bar strength amplitude $A_2/A_0$ is $60 \%$ of maximum value, and this radius lies beyond the peak bar strength. In model D3 evolution, disk loses the maximum angular momentum within the bar length region while the inner halo loses lesser angular momentum. This explains the growth of the bar in model D3,  while the suppression of the bar in model D0 is seen in the left panel of Figure \ref{fig:BS_PS}. We find that the outer disk and halo (radius $>$ bar length) gain angular momentum for all the cases. This angular momentum gain in outer disks becomes larger for models having lesser halo angular momentum distribution discontinuity. 

\begin{figure*}
    \centering
    \includegraphics[scale=0.5]{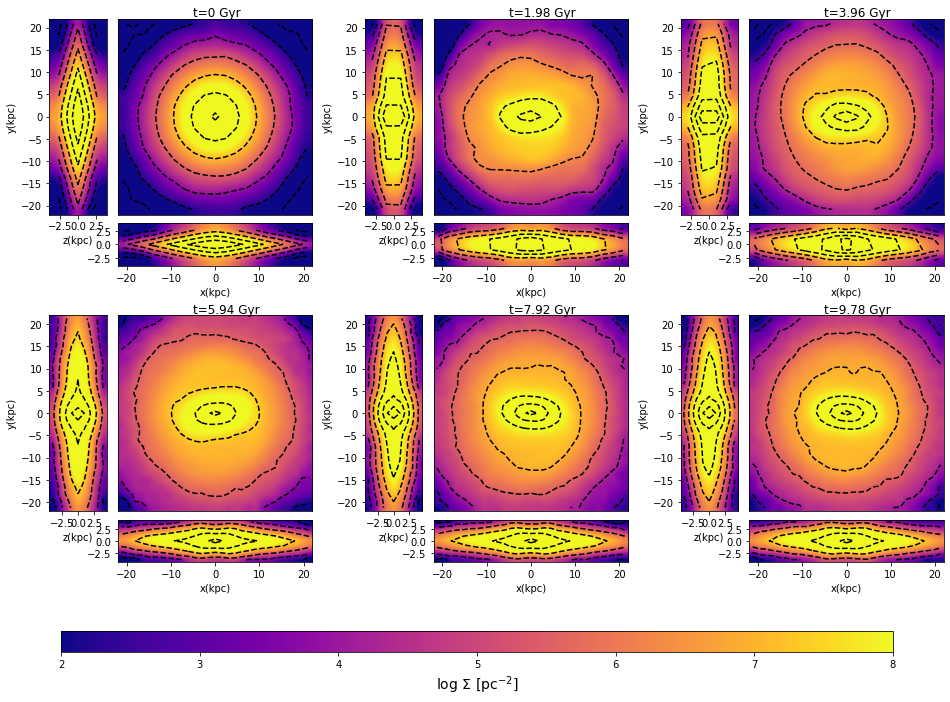}
    \caption{The time evolution of face-on and edge-on density maps of model D0. This shows that the growth of the bar is suppressed in the secular evolution phase, and bar strength doesn't grow.}
    \label{fig:faceon_time_D0}
\end{figure*}

\begin{figure*}
    \centering
    \includegraphics[scale=0.5]{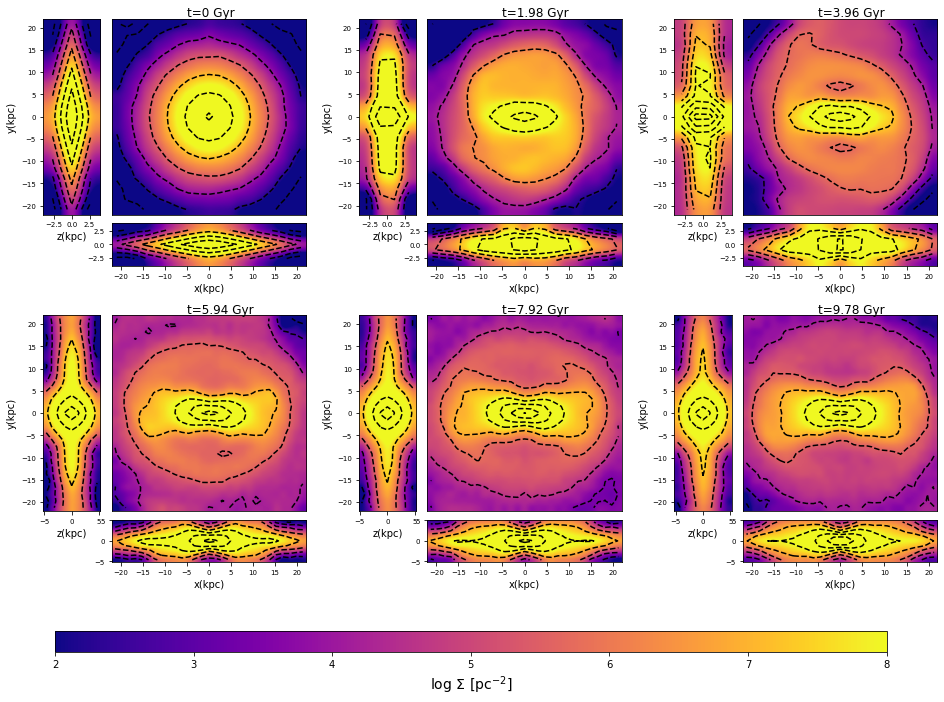}
    \caption{The time evolution of face-on and edge-on density maps of model D3. This shows that the bar grows with time as well as gains boxy/peanut/x-shape in edge-on view.}
    \label{fig:faceon_time_D3}
\end{figure*}

\begin{figure*}

    %\centering
    \begin{tabular}{c|c|c|c}
       
       \includegraphics[scale=0.25]{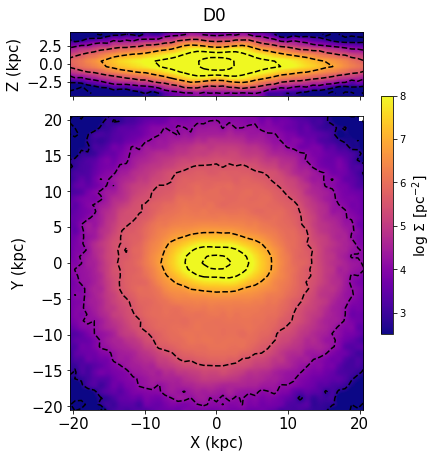}  &
       \includegraphics[scale=0.25]{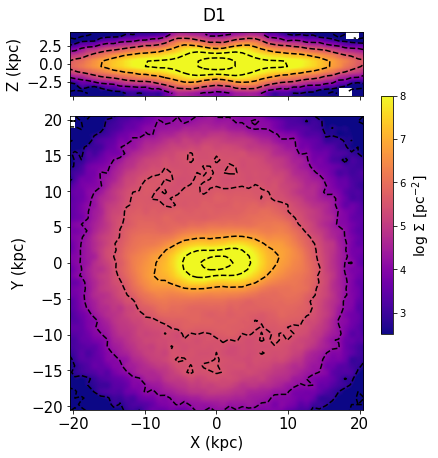}  & \includegraphics[scale=0.25]{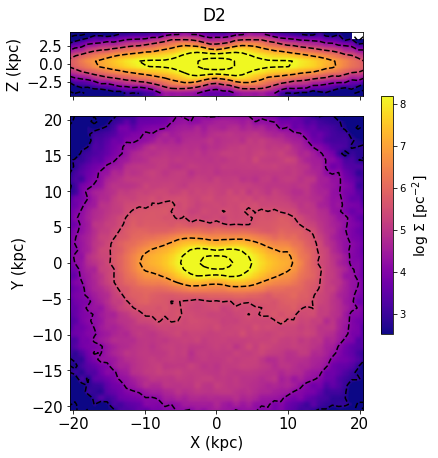} &
       \includegraphics[scale=0.25]{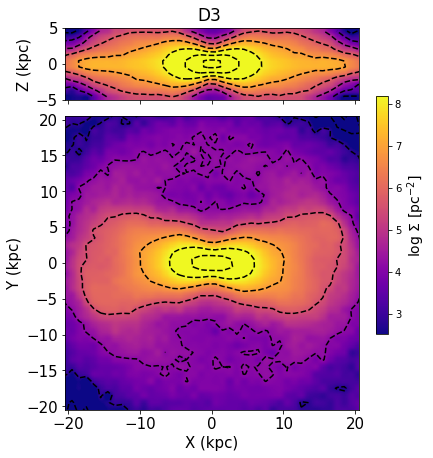}
       
    \end{tabular}
   
    \caption{Face-on and edge-on maps at 9.78 Gyr in models with increasing continuity of halo angular momentum distribution which is lowest in model D0 and highest in model D3. The bar strength and boxy/peanut/x-shape become more prominent as we move from model D0 to model D3.}
 
    \label{fig:faceon_10gyr_all}
\end{figure*}

Figure \ref{fig:faceon_time_D0} shows the time evolution of face-on and edge-on maps of model D0. The bar appears at 1.98 Gyr of evolution, along with a weak spiral along the edges of the bar. We can see that the bar doesn't grow during its evolution and remains weaker until the evolution of 9.78 Gyr, as also seen in the left panel of Figure \ref{fig:BS_PS} of bar strength evolution with time. The edge-on view shows a weak boxy bulge at the end of the simulations around 9.78 Gyr. Figure \ref{fig:faceon_time_D3} shows that the bar in model D3 grows with time. At 3.96 Gyr, we see that ring at the bar's edge, which dissolves as the bar evolves.  We can also see the boxy/peanut/x-shaped bulge growth in the central region, which is more prominent than in model D0. Finally, Figure \ref{fig:faceon_10gyr_all} shows the face-on and edge-on view of all four models D0, D1, D2, and D3, at the end of the simulation. As the continuity of halo angular momentum distribution increases from model D0 to model D3, the secularly evolved bar has the strongest bar strength. The face-on shape of the bar changes from oval in weaker bars to butterfly shape in the strongest bar case. We also find that x-shaped bulges are more prominent in model D3 than in model D0. 

We have measured buckling amplitude's time evolution, given by the following equation \citep{Debattista.etal.2006}.

\begin{equation}
    A_{\text{buckle}}= \left | \dfrac{\sum_{i=1}^{N} z_im_i e^{2i\phi_{i}} }{\sum_{i=1}^{N} m_i} \right |
    \label{eq:Buckle}
\end{equation}

\begin{figure*}
    \centering
    \includegraphics[scale=.7]{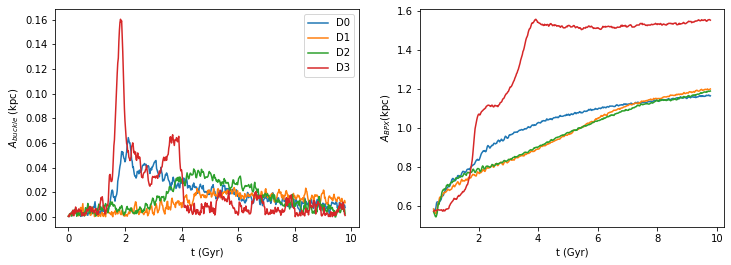}
    \caption{Left and right panels show the time evolution of buckling strength and box/peanut/x-shaped bulge strength, respectively, in all the models}
    \label{fig:buckle_bpx}
\end{figure*}

where $z$ is the vertical coordinate, $m_i$ is mass of $i^{th}$ star, $\phi_i$ is the azimuthal angle.  The left panel of Figure \ref{fig:buckle_bpx} shows the buckling amplitude for all the simulated models. We can see that models D1 and D2 show weak buckling. Model D3 shows the multiple buckling events at 1.9 Gyr and 3.9 Gyr respectively. The amplitude of first buckling is more than twice of the amplitude in second buckling event. Model D0 shows a single buckling having lower amplitude than the first buckling in D3 model. Though the time span of the buckling in D0 is much larger than both buckling events in the model D3.

 We quantify the boxy/peanut/x-shape amplitude ($A_{\text{BPX}}$) of the bar, which is the mean square root of the vertical height of the bar and measured within an annular region of radius ranging from 2 kpc to 8 kpc. We define the buckling strength \citep{Baba.etal.2021}. 
\begin{equation}
    A_{\text{BPX}}= \sqrt{\dfrac{\sum_{k=1}^{N} m_k z_k^2 }{\sum_{k=1}^{N} m_k}} 
    \label{eq:BPX}
\end{equation}

where $z$ is the vertical coordinate, $m_k$ is mass of $k^{th}$ star. The right panel of Figure \ref{fig:buckle_bpx} shows the evolution of $A_{\text{BPX}}$ with time for all the models. We can see that $A_{\text{BPX}}$ increases with time for all the models. A sudden increase in $A_{\text{BPX}}$ at around 1.9 Gyr and 3.9 Gyr corresponds to the first and second buckling events for model D3. The final boxy/peanut/x-shape amplitude strength is highest for model D3 and lowest for models D0, suggesting the huge impact of the halo angular momentum distribution discontinuity.

\section{Discussion}\label{Discussion}
\cite{Kataria.Shen.2022} highlighted in their high-spinning model (S100 of their paper) that the fate of bar growth in the secular evolution phase depends on the angular momentum distribution in the central region. The models used in this article are motivated by a detailed investigation of the role of this discontinuity on the distribution of dark matter angular momentum. The angular momentum transfer from the disk to the halo, crucial to bar growth in secular evolution \citep{Athanassoula.2003}, depends on the discontinuity of halo angular momentum distribution. Figure \ref{fig:AM_exchange} shows that the halo with the most continuous angular momentum distribution (model D3) lose lower inner halo angular momentum than the halo with discontinuous angular momentum distribution (model D0). While the inner disk lose the most of the angular momentum for model D3 compare to model D0. This indicates that the halo particles with a similar distribution of prograde and retrograde motion, basically the similar distribution of positive and negative angular momentum distribution, in the central region lead to a significant exchange of angular momentum from the disk.

To juxtapose the novel aspects of the results of this article, we restate that the halo properties like mass, concentration, and spin \citep{Athanassoula.Misiriotis.2002,Collieretal.2018,Kataria.Shen.2022} play an important role in the formation and evolution of bar-type instabilities. In the present study, we highlight that for a given total angular momentum, mass, concentration, and spin ($\lambda$) of dark matter halo, the distribution of angular momentum in the central region also plays a vital role in the secular growth of the bar. The effect of halo angular momentum distribution discontinuity gets mitigated as we go towards the lower spin cases. For the non-rotating halo ($\lambda$=0), we find similar secular evolution irrespective of discontinuity. In literature \citep{Longetal.2014, Collieretal.2019, joshi.widrow.2023}, it is a common practice to increase the halo spin in a simple manner (constant fraction of retrograde orbits at all radii), which introduces the discontinuity of halo angular momentum distribution at $L_z=0$. In reality, the cosmological simulations have a continuous distribution of angular momentum distribution for dark matter halos. These results provide a plausible explanation for conflicting results in studies claiming secular growth of bar \citep{Kataria.Shen.2022} and secular dampening of bar \citep{Longetal.2014,Collieretal.2018} in high spin models.

\section{Conclusions} \label{Conclusion}
In this article, we report the role of dark matter halo angular momentum distribution discontinuity on the bar formation and evolution for the first time in a non-linear regime. This provides insight into the central asymmetric structure of the Milky Way-type galaxies. Our main findings are listed below.

\begin{itemize}
    \item We find that the model (D0) with maximum halo angular momentum distribution discontinuity around $L_z=0$ does not show the bar growth in secular evolution. On the other hand, the model (D3) with the most continuous halo angular momentum distribution shows the highest bar growth in the secular evolution.
    \item The decrease in pattern speed is the highest for the model having the most continuous halo angular momentum distribution and vice versa. 
    \item The angular momentum transfer from the inner disk (within the bar length region) to the outer disk (outside the bar length region) is pronounced in the most continuous angular momentum distribution cases.
    \item The buckling events are more prominent in the most continuous distribution of halo angular momentum. This results in the most pronounced boxy bulges in models with continuous angular momentum distribution.

\end{itemize}

These results provide insights into the role of unnatural initial conditions in the N-body community, specifically the severe effect of the discontinuous halo angular momentum distribution.

\section*{Acknowledgements}

We would like to thank the anonymous referee for the insightful report which has significantly improved the work. We thank Jerry Sellwood, Victor Debattista and Rimpei Chiba for helpful discussions on the discontinuity of dark matter angular momentum distribution. We also thank Sioree Ansar for the helpful discussions on dark matter halo spin. The research presented here is partially supported by the National Key R\&D Program of China under grant No. 2018YFA0404501; by the National Natural Science Foundation of China under grant Nos. 12025302, 11773052, 11761131016; by the ``111'' Project of the Ministry of Education of China under grant No. B20019; and by the China Manned Space Project under grant No. CMS-CSST-2021-B03. J.S. also acknowledges support from a \textit{Newton Advanced Fellowship} awarded by the Royal Society and the Newton Fund. We thank Volker Springel for the GADGET code we used to run our simulations. This work used the Gravity Supercomputer at the Department of Astronomy, Shanghai Jiao Tong University, and the Center for High-Performance Computing facilities at Shanghai Astronomical Observatory. 
software used: numpy \citep{Harris.et.al.2020}, matplotlib \citep{Hunter.2007}, pynbody \citep{Pynbody.2013} and astropy \citep{Astrop.collaboration.2018} packages. 

%%%%%%%%%%%%%%%%%%%%%%%%%%%%%%%%%%%%%%%%%%%%%%%%%%
\section*{Data Availability}

We provide the key snapshots of our all models on "https://zenodo.org/records/11356883".

%%%%%%%%%%%%%%%%%%%% REFERENCES %%%%%%%%%%%%%%%%%%

% The best way to enter references is to use BibTeX:

\bibliographystyle{aasjournal}
\bibliography{example} % if your bibtex file is called example.bib

% Alternatively you could enter them by hand, like this:
% This method is tedious and prone to error if you have lots of references
%\begin{thebibliography}{99}
%\bibitem[\protect\citeauthoryear{Author}{2012}]{Author2012}
%Author A.~N., 2013, Journal of Improbable Astronomy, 1, 1
%\bibitem[\protect\citeauthoryear{Others}{2013}]{Others2013}
%Others S., 2012, Journal of Interesting Stuff, 17, 198
%\end{thebibliography}

%%%%%%%%%%%%%%%%%%%%%%%%%%%%%%%%%%%%%%%%%%%%%%%%%%

%%%%%%%%%%%%%%%%% APPENDICES %%%%%%%%%%%%%%%%%%%%%
\appendix

\section{Test with the impact of the orbit reversal with upper limit on $L_{z}$} \label{appendix}
\begin{figure}
    \centering
    \includegraphics[scale=0.65]{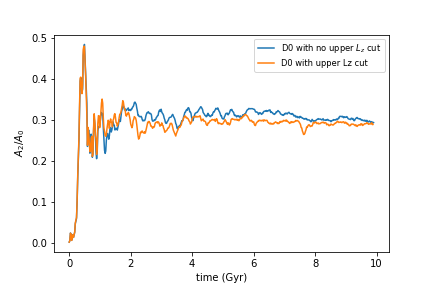}
    \caption{The orbital reversal with probability $x>g(L_z)$ of D0 with the model with and without an upper limit on Lz leads to a similar evolution in the bar strength.}
    \label{fig:Test_UC}
\end{figure}

We have conducted a test of imposing the upper limit on angular momentum while flipping the particles for model D0. We show that models D0 with an upper limit on orbital reversal (spin $= 0.1$) and model D0 without an upper limit on orbital reversal (spin $>>0.1$) lead to a similar evolution in the growth of bar evolution. Figure \ref{fig:Test_UC} shows that both models form bars at similar times with similar bar strengths. The bar strength evolution is almost similar during the secular evolution and the final bar strength is the same for both models. Hence, we confirm that imposing an upper limit does not impact the evolution of galaxy models.

%%%%%%%%%%%%%%%%%%%%%%%%%%%%%%%%%%%%%%%%%%%%%%%%%%

% Don't change these lines
%\bsp	% typesetting comment
%\label{lastpage}
%\end{document}

% End of mnras_template.tex

%\appendix

%\section{Face-on and edge-on time evolution of models}
%Figure \ref{fig:faceon_time_D0} and \ref{fig:faceon_time_D3} show the time evolution of face-on and edge-on density maps of model D0 and D3, respectively.

% Don't change these lines
%\bsp	% typesetting comment
\label{lastpage}
\end{document}